\documentclass[10pt, conference, letterpaper]{IEEEtran}
\usepackage{cuted}
\usepackage{listings}
\usepackage{amsfonts}

\usepackage{multicol}

\usepackage{blindtext}
\usepackage{amsfonts}
\usepackage{amsmath, amsthm, amssymb}
\usepackage{times}
\usepackage{url}
\usepackage{verbatim}
\usepackage{array}
\usepackage{mathrsfs}
\usepackage{enumitem}
\usepackage{textcomp}

\newcommand{\subparagraph}{}
\usepackage[compact]{titlesec}
\usepackage[ruled, lined, linesnumbered]{algorithm2e}
\SetAlFnt{\small}

\SetCommentSty{mycommfont}

\usepackage{amsfonts}
\usepackage{bbm}
\usepackage{soul}
\usepackage{color}
\usepackage{graphics}
\usepackage{graphicx,epsfig}
\usepackage{lipsum,graphicx,subcaption}
\graphicspath{{figures/}}
\usepackage{caption}
\usepackage{subcaption}

\usepackage{cleveref}
\usepackage{wrapfig}
\usepackage{url}
\usepackage{url}
\usepackage{amsmath,amssymb,epsfig,amsthm,psfrag,epsf, multirow,wrapfig, graphicx}
\usepackage[english]{babel}
\usepackage{epstopdf}
\usepackage{float}
\usepackage{color}
\usepackage[table,xcdraw]{xcolor}
\usepackage{amsmath}
\usepackage{amsfonts}
\usepackage{amssymb}
\usepackage{mathtools}

\usepackage{tabularx}
\usepackage{adjustbox}
\usepackage{physics}

\usepackage{amsfonts}
\usepackage{amssymb}
\usepackage{caption}
\usepackage{subcaption}
\usepackage{tabularx}
\usepackage{comment}
\usepackage[mathscr]{euscript}
\usepackage{amsbsy}
\usepackage{stmaryrd}
\usepackage{mathtools}
\usepackage[utf8]{inputenc}
\graphicspath{{figs/}}

\usepackage{booktabs}
\usepackage{graphicx,epsfig}
\usepackage{graphics}
\usepackage{multirow}
\usepackage{rotating}
\usepackage{caption}
\usepackage{subcaption}
\usepackage{amsmath}
\usepackage{amssymb}
\usepackage{makecell}

\usepackage{mathtools}
\usepackage{comment}
\usepackage{multirow}

\usepackage{footnote}
\usepackage{tablefootnote}
\usepackage{scalerel}

\usepackage{tabularx}
\usepackage{mdframed}
\usepackage{lipsum}
\makeatletter
\newcommand*{\rom}[1]{\expandafter\@slowromancap\romannumeral #1@}
\makeatother
\usepackage{nccmath}

\usepackage{relsize}

\newtheorem{theorem}{Theorem}

\newtheorem{corollary}{Corollary}

\newtheorem{definition}{Definition}

\newtheorem{lemma}{Lemma}

\renewcommand\IEEEkeywordsname{Keywords}

\DeclareMathSymbol{\shortminus}{\mathbin}{AMSa}{"39}

\allowdisplaybreaks
\begin{document}
\def\eg{\mbox{\em e.g.}, }


\title{Multidimensional Eigenwave Multiplexing Modulation for Non-Stationary Channels}

\author{
\IEEEauthorblockN{Zhibin Zou, and Aveek Dutta}
    \IEEEauthorblockA{Department of Electrical and Computer Engineering\\
    University at Albany SUNY, Albany, NY 12222 USA\\
    \{zzou2, adutta\}@albany.edu}
    \vspace{-5ex}

}
    
\maketitle

\begin{abstract}
OFDM modulation and OTFS modulation have demonstrated their efficacy in mitigating interference in the time and frequency domains, respectively, caused by path delay and Doppler shifts. However, no established modulation technique exists to address inter-Doppler interference (IDI) resulting from time-varying Doppler shifts. Additionally, both OFDM and OTFS require supplementary precoding techniques to mitigate inter-user interference (IUI) in MU-MIMO channels. To address these limitations, we present a generalized modulation method for any multidimensional channel, based on Higher Order Mercer's Theorem (HOGMT)~\cite{ZouICC22, ZouTCOM23} which has been proposed recently to decompose multi-user non-stationary channels into independent fading subchannels (Eigenwaves). The proposed method, called multidimensional Eigenwaves Multiplexing (MEM) modulation, uses jointly orthogonal eigenwaves decomposed from the multidimensional channel as subcarriers, thereby avoiding interference from other symbols transmitted over multidimensional channels. We show that MEM modulation achieves diversity gain in eigenspace, which in turn achieves the total diversity gain across each degree of freedom(\eg space (users/antennas), time-frequency and delay-Doppler). The accuracy and generality of MEM modulation are validated through simulation studies on three non-stationary channels.
\end{abstract}
\section{Introduction}
\label{sec:intro}
\begin{figure}
    \centering
    \includegraphics[width=\linewidth]{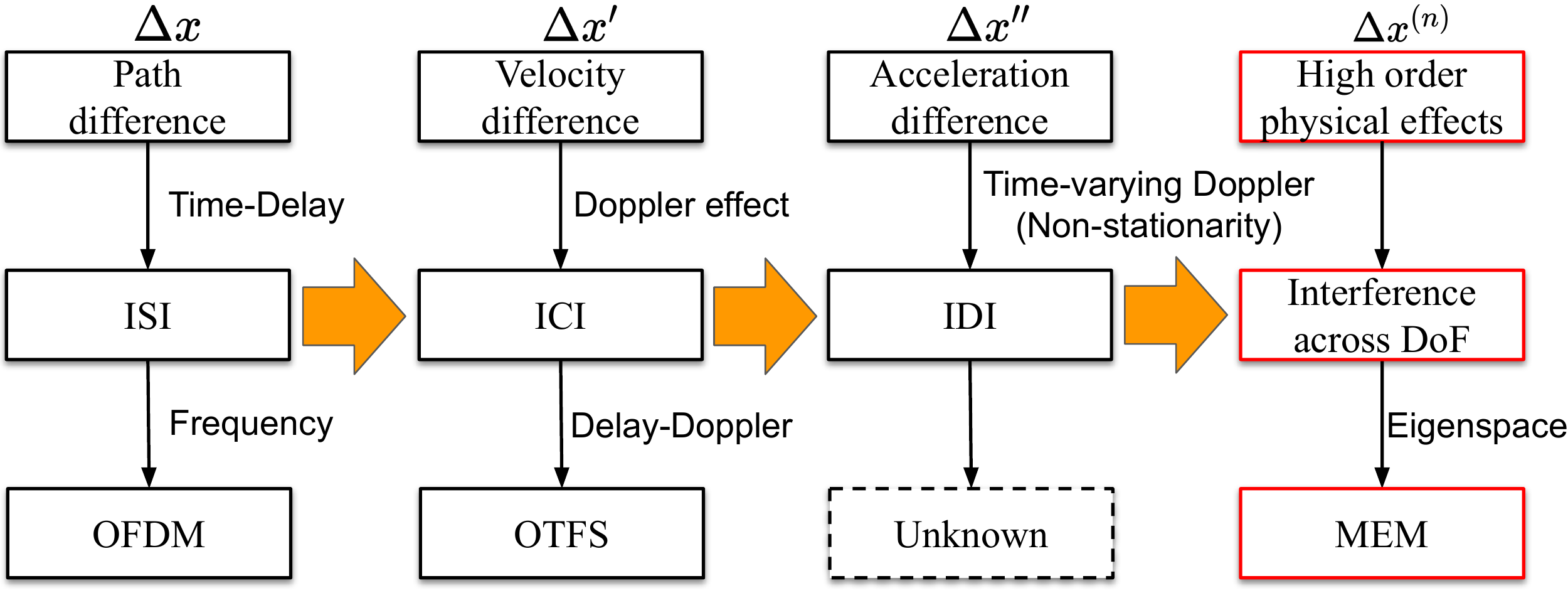}
    \caption{The evolution of modulation techniques}
    \label{fig:evo}
    \vspace{-10pt}
\end{figure}

Path delays cause the Inter-Symbol Interference (ISI), which can be mitigated by OFDM as it transmits symbols in frequency domain~\cite{tse2004}. On the other hand, Doppler effect causes Inter-Carrier Interference (ICI), which can be mitigated by OTFS modulation by transmitting symbols in the delay-Doppler domain~\cite{OTFS_2018_Paper}
However, in non-stationary channels, both the delay and the Doppler effects change over time and frequency, leading to interference at the delay-Doppler domain, which is also referred as Inter-Doppler Interference (IDI)~\cite{Raviteja2018eq_otfs}. Detectors have been investigated to mitigate IDI for OTFS symbols~\cite{Raviteja2018eq_otfs, Thaj2021RxOTFS}. However, these additional techniques can not ensure interference-free at the delay-Doppler domain, especially for highly non-stationary channels. As shown in figure~\ref{fig:evo}, these techniques are developed iteratively to mitigate ISI, ICI and then IDI due to \textit{path difference ($\Delta x$), velocity difference ($\Delta x^{\prime}$) and acceleration difference ($\Delta x^{\prime\prime}$)}, by investigating orthogonality in the time, time-frequency and delay-Doppler domain, respectively. In general, modulation techniques design carriers in the domain represented by high order physics as it is relatively less variant with minimal interference.
This motivates us to investigate a general modulation for high dimensional channels. Moreover, the above modulations can not directly incorporate space domain, therefore requiring additional precoding techniques to cancel spatial interference for MU-MIMO channels~\cite{cao2021low, pandey2021low}. In this paper, we present a general high dimensional modulation for non-stationary channels. Notice Any channel can be generated as a special case of the non-stationary channel. Therefore the proposed modulation for non-stationary channels will certainly generalize to any and all wireless channels. 
Recently, HOGMT has been proposed as a mathematical tool for multi-user non-stationary channel decomposition~\cite{ZouICC22, ZouTCOM23}. It can decompose the high-dimensional channels into independent subchannels along each degree of freedom (DoF). We leverage this tool to develop Multidimensional Eigenwave Multiplexing (MEM) modulation which uses the jointly orthogonal eigenwaves decomposed from the high dimensional channel as carriers. Symbols on these carriers achieve orthogonality across each DoF, thereby avoiding interference from all DoF. 

We summarize the qualitative differences between OTFS and MEM in Table~\ref{tab:cha}. OTFS obtains the carriers by Symplectic Fourier Transform (SFT), while MEM obtains its carriers by HOGMT. The OTFS carriers are in the delay-Doppler domain, while carriers of MEM are at the eigen domain. Further, OTFS requires CSI at the receiver side only, while MEM requires CSI at both transmitter and receiver side. This is the main cost of MEM, although the CSI is generally required at the transmitter in modern wireless systems that is well documented in the literature~\cite{Guo2022CSI}. The OTFS input-output relation and the corresponding modulation schemes for non-stationary channels does not exist currently and it cannot directly generalize to higher dimensional channels. For instance, it requires additional precoding for MU-MIMO channels as it does not achieve spatial orthogonality. The details about the limitations of OTFS are discussed in Section~\ref{sec:Limitations}. The proposed MEM is able to achieve orthogonality for non-stationary channels and generalize to higher dimension, meaning it doesn't require additional detector and precoding to cancel IDI and IUI.   
\setlength{\tabcolsep}{0.2em}
\begin{table}[t]
\vspace{0.03in}
\caption{Comparison between OTFS and MEM}
\renewcommand*{\arraystretch}{1.2}
\centering
\begin{tabular}{|c|c|c|}
\hline
\textbf{Modulations} & \textbf{OTFS} & \textbf{MEM} \\ \hline
Mathematical tool & SFT  &  HOGMT \\ \hline
Carriers domain & Delay-Doppler domain  &  Eigen domain \\ \hline
CSI requirement & CSI at Rx & CSI at Tx and Rx            \\ \hline
Adaptive to NS channels & No & Yes   \\ \hline
General to HD channels & No & Yes \\ \hline
\end{tabular}
\label{tab:cha}
\vspace{-10pt}
\end{table}
To the best of our knowledge, MEM is the first generalized modulation for high-dimensional non-stationary channels.
The contribution of this paper is summarized as follows:
\begin{itemize}
    \item We deduce the input-output relation for non-stationary channels and show the limitations of OTFS modulation, which is susceptible to IDI due to OTFS symbols being unable to maintain orthogonality for time- and frequency-varying delay-Doppler response.
    \item We design a multidimensional 
    modulation with 
    jointly orthogonal eigenwaves as subcarriers,  
    cancelling interference in all degrees of freedom in 
    non-stationary channels.
    \item We show the generality of MEM to stationary channels, non-stationary channels and higher dimensional channels, where we validate higher dimensional generality by extending the non-stationary channel to space domain (\eg MU-MIMO non-stationary channels) and demonstrate MEM can also cancel spatial interference without additional precoding. 
    \item We validate MEM under three channels, two of which show the performance in non-stationary channels with different non-stationarity intervals. The third shows its generality by incorporating the spatial domain.
\end{itemize}

\section{Preliminary}
\label{sec:pre}

\begin{figure}
    \centering
    \includegraphics[width=\linewidth]{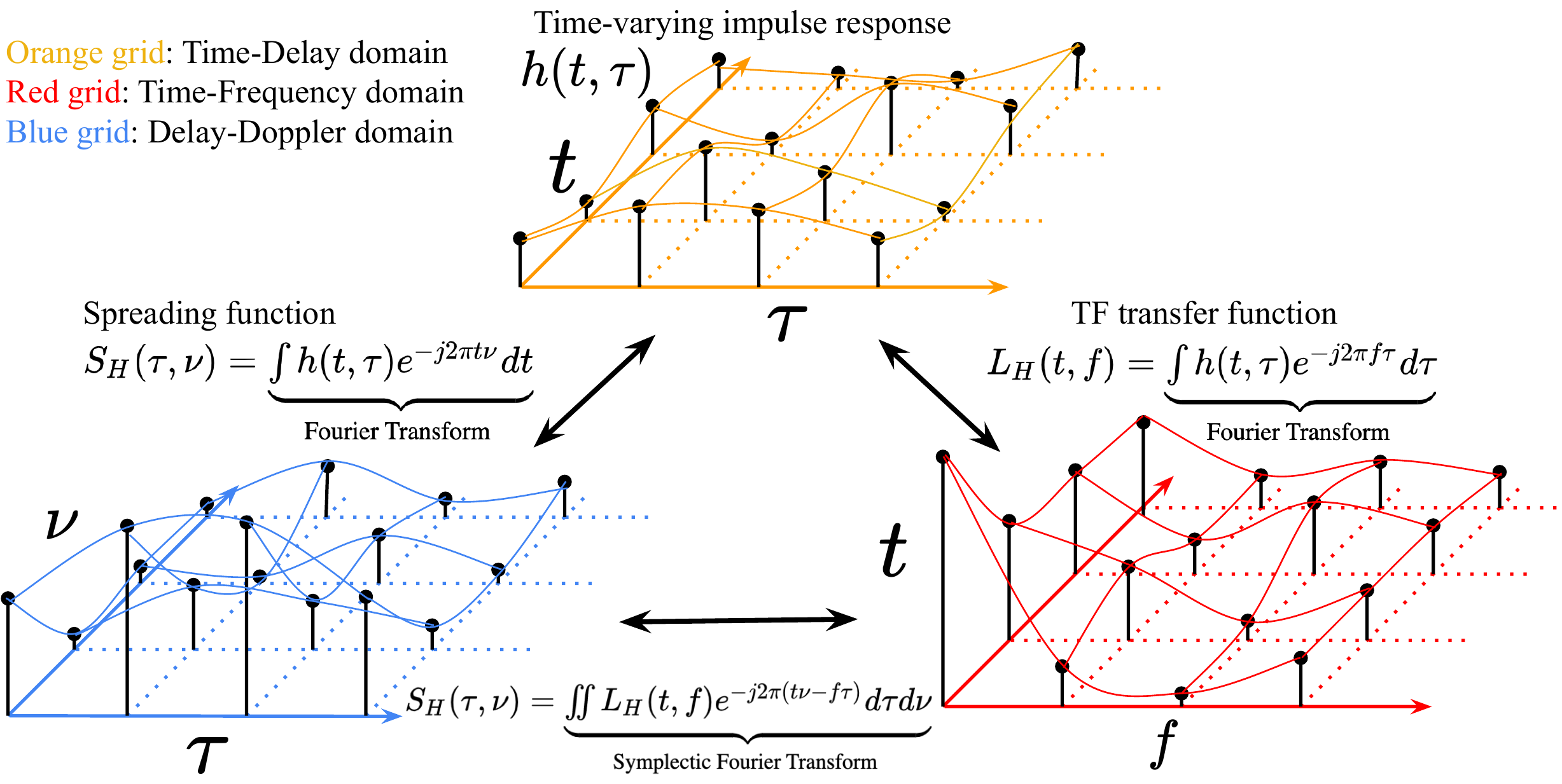}
    \caption{General LTV model transition in 4-D domain (time, frequency, delay and Doppler)~\cite{MATZ20111,Bello1963Chara}.}
    \label{fig:4-D_relation}
    \vspace{-10pt}
\end{figure}

The wireless channel is typically expressed by a linear operator $H$, and the received signal $r(t)$ is given by $r(t){=}Hs(t)$, where $s(t)$ is the transmitted signal. The physics of the impact of $H$ on $s(t)$ is described using the delays and Doppler shift in the multipath propagation~\cite{MATZ20111} given by \eqref{eq:H_delay_Doppler},
\begin{equation}
    r(t) = \sum\nolimits_{p=1}^P h_p s(t-\tau_p) e^{j2\pi \nu_p t}
    \label{eq:H_delay_Doppler}
\end{equation}
where $h_p$, $\tau_p$ and $\nu_p$ are the path attenuation factor, time delay and Doppler shift for path $p$, respectively. 
\eqref{eq:H_delay_Doppler} is expressed in terms of the overall delay $\tau$ and Doppler shift $\nu$ \cite{MATZ20111} in \eqref{eq:S_H}, 
%
\begin{align}
    &r(t) = \iint S_H(\tau, \nu) s(t{-} \tau) e^{j2\pi \nu t} ~d\tau ~d\nu \label{eq:S_H} \\
    & = \int L_H(t, f) S(f) e^{j2\pi tf} ~df 
     = \int h(t, \tau) s(t{-} \tau) ~d\tau \label{eq:h_relation}
\end{align}
where $S_H(\tau, \nu)$ is the \textit{(delay-Doppler) spreading function} of channel $H$, which describes the combined attenuation factor for all paths in the delay-Doppler domain. $S(f)$ is the Fourier transform of $s(t)$ and the time-frequency (TF) domain representation of $H$ is characterized by its \textit{TF transfer function}, $L_H(t,f)$ which can be obtained by 2-D Fourier transform as~\eqref{eq:TF_SH}. The time-varying impulse response $h(t,\tau)$ is obtained as the Inverse Fourier transform of $S_H(\tau, \nu)$ from the Doppler domain to the time domain as in \eqref{eq:h_SH}.
\begin{align}
    &L_H(t,f) {=} \iint S_H(\tau, \nu) e^{j2\pi (t\nu{-} f \tau)} ~d\tau ~d\nu \label{eq:TF_SH}\\
    &h(t,\tau) {=} \int S_H(\tau, \nu) e^{j2\pi t \nu} ~d\nu \label{eq:h_SH}
\end{align}
Figure~\ref{fig:4-D_relation} shows a general Linear Time Varying (LTV) channel model, represented in different domains and illustrates the mutual relationship between $h(t, \tau)$, $L_H(t,f)$ and $S_H(\tau, \nu)$. For non-stationary channels, the second-order statistics are 4-D functions. The details are given in Appendix~A~\cite{ICC23appendix}.
%

\section{Limitations of OTFS}
\label{sec:Limitations}

\noindent
\textbf{OTFS input-output relation}: The OTFS delay-Doppler input-output relation~\cite{OTFS_2018_Paper} can be rewritten in continuous form as,
\begin{equation}
\label{eq:otfs_1}
    r(t,f) = \iint h_w(\tau, \nu) s(t - \tau, f - \nu) ~d\tau ~d\nu + v(t,f)
\end{equation}
where $v(\tau,\nu)$ is noise, and $h_w (\tau, \nu)$ is the twisted convolution of delay-Doppler response, $h_c(\tau, \nu)$ with window function $w(\tau,\nu)$ (Heisenberg transform) as in \eqref{eq:otfs_2},
\begin{equation}
\label{eq:otfs_2}
h_{w}(\tau, \nu){=}\iint e^{-j 2 \pi \nu^{\prime} \tau^{\prime}} h_c\left(\tau^{\prime}, \nu^{\prime}\right) w\left(\nu{-}\nu^{\prime}, \tau{-}\tau^{\prime}\right) \mathrm{d} \tau^{\prime} \mathrm{d} \nu^{\prime} 
\end{equation}
and relation of spreading function $S_H(\tau,\nu)$ and delay-Doppler response $h_c(\tau,\nu)$, as given in~\cite{Bello1963Chara} is,
\begin{equation}
    S_H(\tau,\nu) =  e^{-j 2 \pi \nu^{\prime}} h_c(\tau, \nu)
\end{equation}
Then~\eqref{eq:otfs_2} is rewritten as 
\begin{equation}
h_{w}(\tau, \nu)=\iint S_H(\tau^{\prime}, \nu^{\prime}) w\left(\nu-\nu^{\prime}, \tau-\tau^{\prime}\right) \mathrm{d} \tau^{\prime} \mathrm{d} \nu^{\prime}
\label{eq:h_w_2d}
\end{equation}
Notice $h_w(\tau, \nu)$ is the flipped correlation of spreading function and a window function, i.e., $h_w(\tau, \nu) = \mathbb{E}\{S_H(\tau^{\prime}, \nu^{\prime}) w(\nu-\nu^{\prime}, \tau-\tau^{\prime})\}$, which implies the stationary channel assumption.\textit{Currently, the model of OTFS input-output relation for non-stationary channels is not available in the literature.}

\noindent
\textbf{Limitations of OTFS in non-stationary channels}:
For non-stationary channels, we have  $\mathbb{E}\{S_H(\tau^{\prime}, \nu^{\prime}) w(\nu{-}\nu^{\prime}, \tau{-}\tau^{\prime})\} = h_w(\tau^{\prime}, \nu^{\prime}; \tau, \nu)$. Then $h_w(\tau, \nu)$ in~\eqref{eq:h_w_2d} is extended to time- and frequency-varying case to $h_w(t,f; \tau, \nu)$, which we define as \textit{local delay-Doppler response} (LDR).

\begin{definition}
(Local delay-Doppler Response) In the non-stationary channels, $h_w(\tau, \nu)$ undergoes a transformation to $h_w(\tau^{\prime}, \nu^{\prime}; \tau, \nu)$, which can subsequently be transferred to $h_w(t,f; \tau, \nu)$ as

\begin{align}
\label{eq:deflsf}
    & h_w(t,f; \tau, \nu) \triangleq \mathbb{F}^2 \{\mathbb{E}\{S_H(\tau^{\prime}, \nu^{\prime})w(\nu-\nu^{\prime}, \tau-\tau^{\prime})\}\} \nonumber \\
    & = \iint h_w(\tau^{\prime}, \nu^{\prime}; \tau, \nu)  \times e^{j2\pi (t \nu^\prime - f \tau^\prime)} ~d\tau^\prime ~d \nu^\prime
\end{align}
where $\mathbb{F}^2$ is the Symplectic Fourier Transform (SFT), i.e., 2-D Fourier Transform. 
\end{definition}

Therefore OTFS input-output relation \eqref{eq:otfs_1} for non-stationary channel is reformulated as,
\begin{equation}
\label{eq:4_d_h_w}
    r(t,f) {=} \iint h_w(t,f; \tau, \nu) s(t - \tau, f - \nu) ~d\tau ~d\nu + v(t,f)
\end{equation}

Note that \eqref{eq:4_d_h_w} has a similar form as \eqref{eq:otfs_1} but shows the time and frequency variation of the impulse response function $h_w(\tau,\nu)$. 
The above deduction shows that OTFS symbols cannot maintain orthogonality in non-stationary channels as the delay-Doppler domain is no longer independent of the time-frequency domain which leads to interference in the delay-Doppler domain. \textit{Currently, the OTFS modulation for non-stationary channels is not available in the literature.}

\noindent
\textbf{Limitations of OTFS in higher dimensional channels}:
Consider the deduced OTFS input-output relation for non-stationary channels in (16). Let $k(t,f;t',f') \triangleq h_w (t,f; t-t', f-f')$ be the channel kernel, then (16) is rewritten as, 
\begin{equation}
\label{eq:h}
    r(t,f) = \iint k (t,f; t', f') s(t', f') ~dt' ~df' + v(t,f)
\end{equation}

For MU-MIMO non-stationary channels, $h_w(t,f;\tau,\nu)$ is extended to $\mathbf{H}(t,f;\tau,\nu)$. For notational convenience, we use $u$ and $u'$ to represent a continuous space domain (users/antennas) at the receiver and transmitter, respectively. Then $\mathbf{H}(t,f;\tau,\nu)$ is henceforth rewritten as $h_w(u,t,f;u',\tau,\nu)$ and thus~\eqref{eq:h} can also be extended to MU-MIMO case as in \eqref{eq:u_k},

\begin{multline}
\label{eq:u_k}
    r(u, t ,f) = \iiint k (u, t,f;u', t', f') s(u', t', f') 
    ~du' ~dt' ~df' \\ + v(u,t,f)
\end{multline}
where $k (u, t,f;u', t', f')  \triangleq h_w (u,t,f;u', t-t', f-f') $ denotes the space-time-frequency transfer function. The OTFS symbol is not able to achieve joint orthogonality at space-time-frequency domain thereby leading to not only the spatial interference but also the joint space-time-frequency interference.



In conclusion, OTFS modulation has the limitations: 1) OTFS cannot deal with non-stationary channels where the delay-Doppler response is time- and frequency-varying. 2) OTFS require additional equalizer for MU-MIMO channels. Moreover, the equalizer and OTFS can only achieve the optimal interference cancellation at space and time-frequency domain respectively, as modulation and equalizers are independent processes. It is not able to achieve the global optimization for the joint space-time-frequency interference cancellation.

\begin{table*}[h]
\vspace{0.09in}
\begin{align}
    & r (\zeta_1{,}{...}{,}\zeta_P) 
=\int{\ldots}\int H(\zeta_1{,}{...}{,}\zeta_P{;} \gamma_1{,}{...}{,} \gamma_Q) s(\gamma_1{,}{...}{,} \gamma_Q) ~d\gamma_1 {,}{...}{,} ~d\gamma_Q {+} v(\zeta_1{,}{...}{,}\zeta_P) \nonumber \\
&=  \int{\ldots}\int \bigg\{ \underbrace{{\sum_{n{=}1}^ N} \sigma_n \psi_n(\zeta_1{,}{...}{,}\zeta_P) \phi_n(\gamma_1{,}{...}{,}\gamma_Q) \sum_n^N s_n \phi_n(\gamma_1,{\ldots}, \gamma_Q) \bigg\}}_{\text{Lemma}~1} ~d\gamma_1 {,}{...}{,} ~d\gamma_Q  + v(\zeta_1{,}{...}{,}\zeta_P) \nonumber \\
&=  \int{\ldots}\int \bigg\{ {\sum_{n{=}1}^ N} \sigma_n s_n \psi_n(\zeta_1{,}{...}{,}\zeta_P) \underbrace{|\phi_n(\gamma_1{,}{...}{,}\gamma_Q)|^2}_{= 1} {+} \sum_{n'\neq n}^N \sigma_n s_{n'} \psi_n(\zeta_1{,}{...}{,}\zeta_P) \underbrace{\phi_n(\gamma_1{,}{...}{,}\gamma_Q) \phi_{n'}(\gamma_1{,}{...}{,}\gamma_Q)}_{= 0} \bigg\} ~d\gamma_1 {,}{...}{,} ~d\gamma_Q + v(\zeta_1{,}{...}{,}\zeta_P) \nonumber \\
&{=} \sum_n^N \sigma_n s_n \psi_n(\zeta_1{,}{...}{,}\zeta_P) + v(\zeta_1{,}{...}{,}\zeta_P) \label{eq:r_n} \tag{17}\\
& \hat{s}_n =\int{\ldots}\int r (\zeta_1{,}{...}{,}\zeta_P) \psi_n^*(\zeta_1{,}{...}{,}\zeta_P) ~d\zeta_1 {,}{...}{,} ~d\zeta_P \nonumber \\ 
& {=} \int{\ldots}\int \sum_n \sigma_n s_n \psi_n(\zeta_1{,}{...}{,}\zeta_P)\psi_n^*(\zeta_1{,}{...}{,}\zeta_P) ~d\zeta_1 {,}{...}{,}  ~d\zeta_P {+} \int{\ldots}\int v(\zeta_1{,}{...}{,}\zeta_P) \psi_n^*(\zeta_1{,}{...}{,}\zeta_P) ~d\zeta_1 {,}{...}{,}  ~d\zeta_P \nonumber \\
& {=} \int{\ldots}\int \sigma_n s_n |\psi_n(\zeta_1{,}{...}{,}\zeta_P)|^2 ~d\zeta_1 {,}{...}{,} ~d\zeta_P + v_n = \sigma_n s_n  + v_n \implies \text{Interference-free data symbols across all degrees of freedom}
\label{eq:s_n} \tag{18}
\end{align}
\hrule
\vspace{-15pt}
\end{table*}

\section{Eigenwave Modulation}
\subsection{HOGMT decomposition - a brief background}

In \cite{ZouICC22}, authors derived a generalized version of Mercer's Theorem~\cite{1909Mercer} for asymmetric kernels and extended it to higher-order kernels, which decomposes an asymmetric multidimensional channel into jointly orthogonal subchannels. 
For any multidimensional process $K(\zeta_1{,}{...}{,}\zeta_P{;} \gamma_1{,}{...}{,} \gamma_Q)$, it can be decomposed by Theorem~1 in ~\cite{ZouICC22} as,
\begin{align}
\label{eq:col}
K(\zeta_1{,}{...}{,}\zeta_P{;} \gamma_1{,}{...}{,} \gamma_Q) {=} {\sum_{n{=}1}^N} \sigma_n \psi_n(\zeta_1{,}{...}{,}\zeta_P) \phi_n(\gamma_1{,}{...}{,}\gamma_Q) \nonumber
\end{align}
where $\mathbb{E}\{\sigma_n \sigma_n'\} {=} \lambda_n \delta_{nn'}$. $\lambda_n$ is the $n$\textsuperscript{th} eigenvalue. $\{\phi_n\}$ and $\{\psi_n\}$ are eigenfunctions having orthonormal property as, 
\begin{equation}
    \begin{aligned}
    &\int{\ldots}\int \phi_n(\gamma_1{,}{...}{,}\gamma_Q) \phi_{n'}(\gamma_1{,}{...}{,}\gamma_Q) ~d\gamma_1 {,}{...}{,} ~d\gamma_Q = \delta_{nn'} \\
    &\int{\ldots}\int \psi_n(\zeta_1{,}{...}{,}\zeta_P) \psi_{n'}(\zeta_1{,}{...}{,}\zeta_P) ~d\zeta_1 {,}{...}{,} ~d\zeta_P = \delta_{nn'} \nonumber
\end{aligned}
\end{equation}

\subsection{Multidimensional Eigenwave Multiplexing modulation}
We leverage the decomposition framework in \cite{ZouICC22} and redefine the eigenfunctions with multiple variables, which can be defined as eigenwaves in multiple dimensions.

\begin{lemma}
(Associative property of eigenwave set projection) Define $\Phi_{a} {=} \Sigma_n^N a_n \phi_n (\gamma_1{,}{...}{,}\gamma_Q)$, we have
\begin{align}
    \langle \Phi_a ,\Phi_b^*\rangle = \langle\Phi_{ab}, \Phi^*\rangle = \langle \Phi , \Phi_{ab}^*\rangle
\end{align}
where $\langle \cdot, \cdot \rangle$ is the eigenwave set projection operator. $\phi_n(\gamma_1{,}{...}{,}\gamma_Q)$ is $Q$ dimensional eigenfunction.
\label{lemma:ass}
\end{lemma}
\begin{proof}
    The proof 
    is provided in Appendix B~\cite{ICC23appendix}. 
    \vspace{-10pt}
\end{proof}


\begin{figure}
    \centering
    \includegraphics[width=\linewidth]{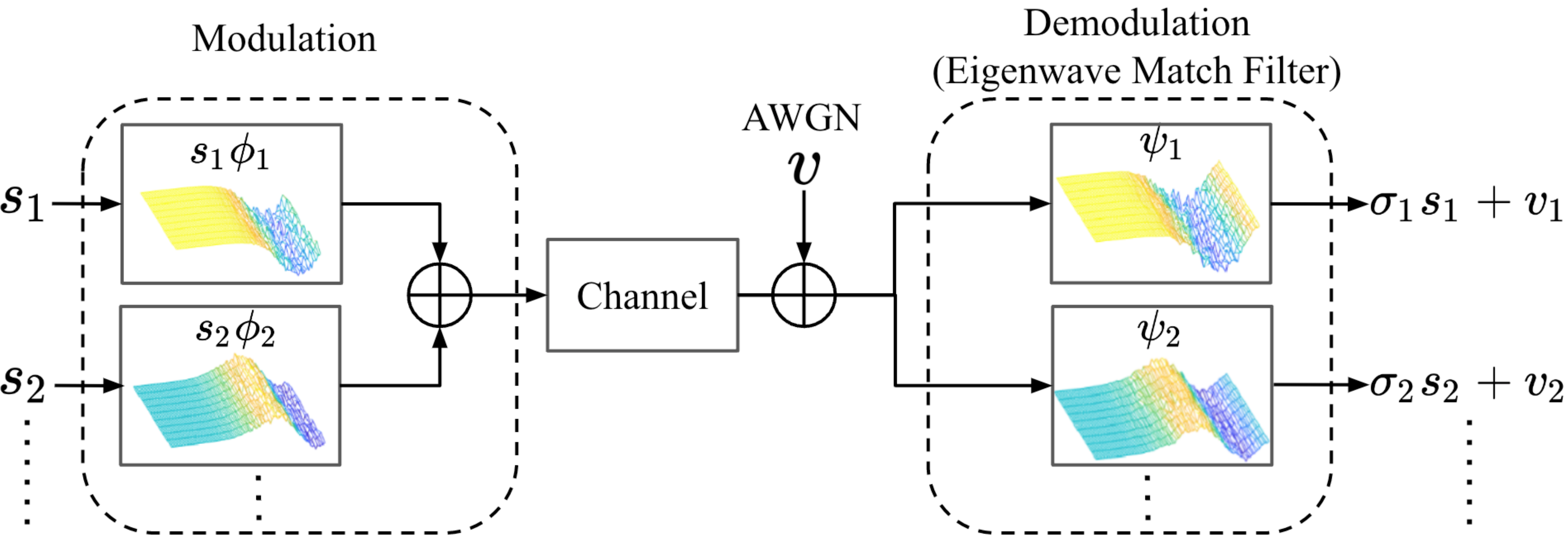}
    \caption{2-D Eigenwave Multiplexing Modulation}
    \label{fig:2-D example}
    \vspace{-10pt}
\end{figure}
\begin{theorem}
\label{thm:mem}
(Multidimensional Eigenwave Multiplexing Modulation and Matched Filter) 

Given, a $M {=} Q{+}P$ dimensional channel transfer function $H(\zeta_1{,}{...}{,}\zeta_P{;} \gamma_1{,}{...}{,} \gamma_Q)$ with input-output relation as
\begin{multline}
    r (\zeta_1{,}{...}{,}\zeta_P) \\=
\int{\ldots}\int H(\zeta_1{,}{...}{,}\zeta_P{;} \gamma_1{,}{...}{,} \gamma_Q) s(\gamma_1{,}{...}{,} \gamma_Q) ~d\gamma_1 {,}{...}{,} ~d\gamma_Q \\ 
+ v(\zeta_1{,}{...}{,}\zeta_P)  \nonumber
\end{multline}
is decomposed into multidimensional eigenfunctions~\cite{ZouICC22} as,
\begin{align}
\label{eq:col}
H(\zeta_1{,}{...}{,} \zeta_P{;} \gamma_1{,}{...}{,} \gamma_Q) {=} {\sum_{n{=}1}^ N} \sigma_n \psi_n(\zeta_1{,}{...}{,}\zeta_P) \phi_n(\gamma_1{,}{...}{,}\gamma_Q) \nonumber
\end{align}
then, a given symbol set, $\{s_n\}$ is modulated using eigenfunctions $\{\phi_n^*\}$ as subcarriers given by,
\begin{align}
    s(\zeta_1{,}{...}{,} \zeta_P) = {\sum_n^N} s_n \phi_n^*(\gamma_1{,}{...}{,} \gamma_Q)
\end{align}
Demodulating the received signal $r (\zeta_1{,}{...}{,}\zeta_P)$ is accomplished by employing the eigenwave matched filter, $\{\psi_n^*\}$ and the estimate $\hat{s}_n$ is given by,
\begin{equation}
    \hat{s}_n = \sigma_n s_n + v_n
\end{equation}
where, $v_n$ is the projection of noise $v(\zeta_1{,}{...}{,}\zeta_P)$ onto the eigenwave $\psi_n^*(\zeta_1{,}{...}{,}\zeta_P)$.  
\end{theorem}
\vspace{-14pt}
\begin{proof}
Transmitting the modulated symbol $s(\gamma_1{,}{...}{,} \gamma_Q)$ over the multidimensional channel with transfer function $H(\zeta_1{,}{...}{,}\zeta_P{;} \gamma_1{,}{...}{,} \gamma_Q)$, the received signal is obtained by~\eqref{eq:r_n}. Demodulating $r (\zeta_1{,}{...}{,}\zeta_P)$ with $\psi_n^*(\zeta_1{,}{...}{,}\zeta_P)$, the estimated data $\hat{s}_n$ is given by~\eqref{eq:s_n}, which suggests that the demodulated symbol $\hat{s}_n$ is the data symbol $s_n$ multiplied a scaling factor (channel gain) $\sigma_n$ along with AWGN, meaning there is no interference from other symbols.
\end{proof}
Figure~\ref{fig:2-D example} shows an example of 2-D eigenwave modulation using Theorem~\ref{thm:mem}. At the transmitter, each data symbol, $s_n$ is multiplied by one eigenwave $\phi_n$, obtained by HOGMT decomposition and then summed to create the modulated signal. The data symbols remain independent during transmission over the channel due to the joint orthogonality of eigenwaves. At the receiver, each each data symbol estimate, $\hat{s}_n$ is obtained by a matching filter using the eigenwave $\psi_n$, also obtained by HOGMT decomposition giving the data symbol $s_n$ multiplied by the corresponding channel gain, $\sigma_n$ with AWGN, $v_n$. Theorem~\ref{thm:mem} is applied to non-stationary channels as follows.
\begin{corollary} (MEM modulation for non-stationary channels) Given the non-stationary channel Local delay-Doppler Response (LDR), $h_w(t,f;\tau,\nu)$ in~\eqref{eq:deflsf} with channel kernel $k(t,f;t',f')$, and the input-output relation in \eqref{eq:h}, 
the data set $\{s_n\}$ is modulated by MEM as
\setcounter{equation}{18}
\begin{align}
    s(t,f) = {\sum_n^N} s_n \phi_n^*(t,f)
\end{align}
At the receiver, interference-free estimated symbol $\hat{s}_n$ is obtained by demodulating the received signal $r(t,f)$ using eigenfunctions $\{\psi_n^*\}$ as
\begin{align}
\label{eq:s_n_tf}
    & \hat{s}_n = \iint r (t,f) \psi_n^*(t,f) ~dt ~df = \sigma_n s_n  {+} v_n
\end{align}
where, $\phi_n(t,f)$ and $\psi_n(t,f)$ are the 2-D eigenwave decomposed from $k(t,f;t',f')$ by HOGMT.
\end{corollary}
\begin{proof}
It follows the same steps and deductions as in the proof of Theorem~\ref{thm:mem}, except for non-stationary channels the multidimensional transfer function $H(\zeta_1{,}{...}{,}\zeta_P{;} \gamma_1{,}{...}{,} \gamma_Q)$ is replaced by the channel kernel $k(t,f;t^{\prime},f')$. 
\end{proof}
Therefore, \eqref{eq:s_n_tf} shows that data symbols are only influenced the channel gain and AWGN, while avoiding interference in non-stationary channel by the use of MEM modulation. Furthermore, MEM modulation can also incorporate additional beamformer such as water filling, MVDR, etc., according to the desired optimization criteria~\cite{ali2017beamforming}. However, beamformers and equalizers are out of the scope of this paper. Meanwhile, replacing $k(t,f;t',f')$ by $k(u,t,f;u',t',f')$ in \eqref{eq:u_k}, MEM modulation can directly incorporate the spatial domain without any modification. It means MEM can be directly applied to MU-MIMO channels without additional precoding. 

\subsection{Generalization}
\noindent
\textbf{Stationary channels}: Assuming the channel is ergodic, as the channel is divided into $N$ independent subchannels (for the non-singular channel matrix/tensor, $N$ is the multiplication of the length of each dimension), the capacity of MEM is the summation capacity of $N$ subchannels. Then the average capacity is given by,
\begin{equation}
    \bar{C} = \max_{\{P_n\}} \frac{1}{T}\sum_n^N \log_2 \big(1 + \frac{P_n |\sigma_n|^2}{N_0}\big)
    \label{eq:C}
\end{equation}
where, $T$ is the time length. $P_n$ and $N_0$ is the power of $s_n$ and $v_n$, respectively.~\eqref{eq:C} shows that, with water filling algorithm, MEM achieves the capacity for stationary channels.

\noindent
\textit{Remark 1:} MEM modulation achieves the sum rate in eigenspace, where eigenwaves are independent subchannels. It also implies achieving the diversity gain in eigenspace. 

\begin{figure}
\vspace{0.07in}
    \centering
    \includegraphics[width=0.875\linewidth]{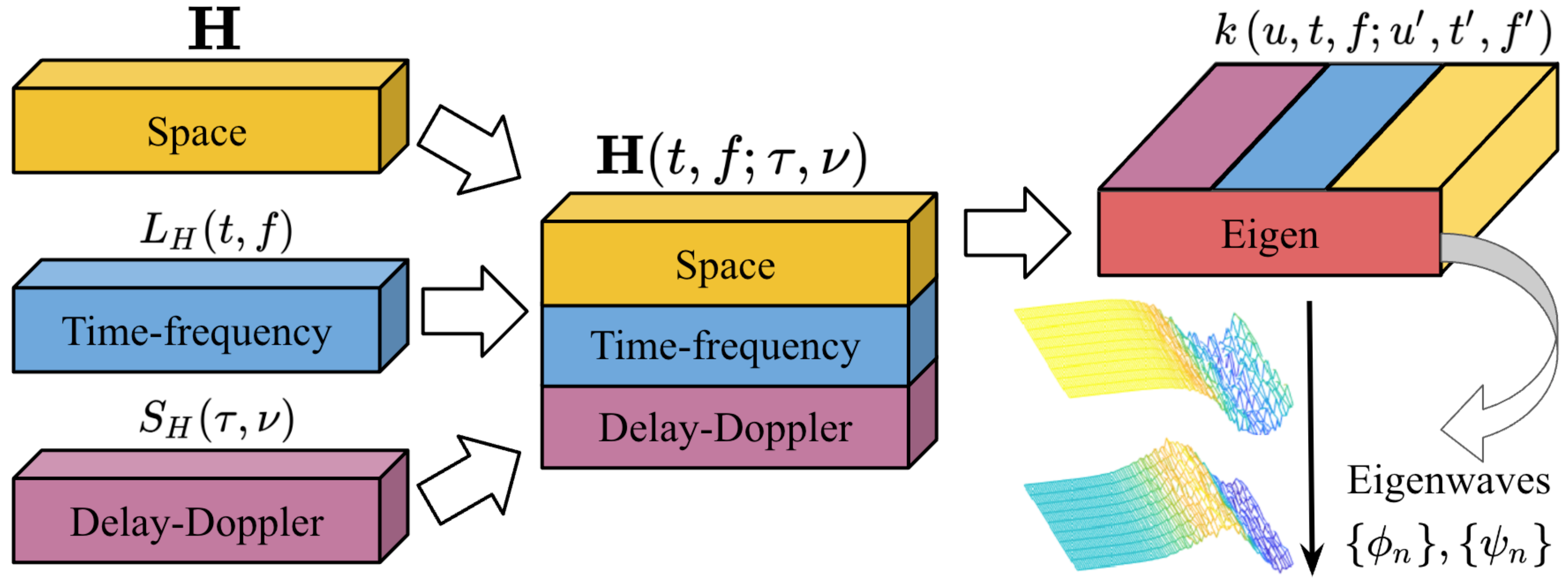}
    \caption{Eigen domain view of space, time-frequency and delay Doppler domain}
    \label{fig:eigendomain}
    \vspace{-15pt}
\end{figure}

\noindent
\textbf{Non-stationary channels}: The capacity for non-stationary channels is not well defined as the ergodic assumption no longer holds. In this case, we give a qualitative analysis about the optimality by using the concept of \textit{``diversity achieving"} for the non-stationary wireless channels. We know from~\cite{ZouICC22} that the total channel gain for the non-stationary channel LDR $h_w(t,f;\tau,\nu)$ in~\eqref{eq:deflsf} is given by,
\begin{equation}
    \iiiint |h_w(t,f;\tau,\nu)|^2~dt~df~d\tau~d\nu = \sum_n^N \lambda_n
    \label{eq:gain}
\end{equation}
where, $\lambda_n$ is $n^\text{th}$ eigenvalue, and $\mathbb{E}\{\sigma_n \sigma_n'\} = \lambda_n \delta_{nn'}$. The deduction of~\eqref{eq:gain} is given in Appendix C~\cite{ICC23appendix}.
Meanwhile, the power over all demodulated symbol $\hat{s}_n$ as in~\eqref{eq:s_n_tf} is, 
\begin{equation}
    \mathbb{E}\left\{\left|\sum_n^N \hat{s}_n\right|^2\right\} = \sum_n^N \lambda_n P_n + N_0
    \label{eq:r}
\end{equation}
From \eqref{eq:gain} and \eqref{eq:r} we find that the data symbol $\{s_n\}$ has leveraged all the diversity gain. 

\noindent
\textbf{Higher dimensional channels:} Replacing $h_w(t,f;\tau,\nu)$ in~\eqref{eq:gain} by $\mathbf{H}(t,f;\tau,\nu)$, MEM modulation can still achieve diversity gain for MU-MIMO non-stationary channels. The reason is that the diversity of the multidimensional channel at each DoF (space, time-frequency, delay-Doppler) are merged (integral along each DoF as in~\eqref{eq:gain}) and then divided in the eigenspace into independently eigewaves as shown in Figure~\ref{fig:eigendomain}. Therefore, eigenwaves achieve diversity in eigenspace, implying that \textit{``diversity achieving"} for the total channel as well.


\section{Results}
\begin{figure}
\begin{subfigure}{.22\textwidth}
  \centering
  \includegraphics[width=1\linewidth]{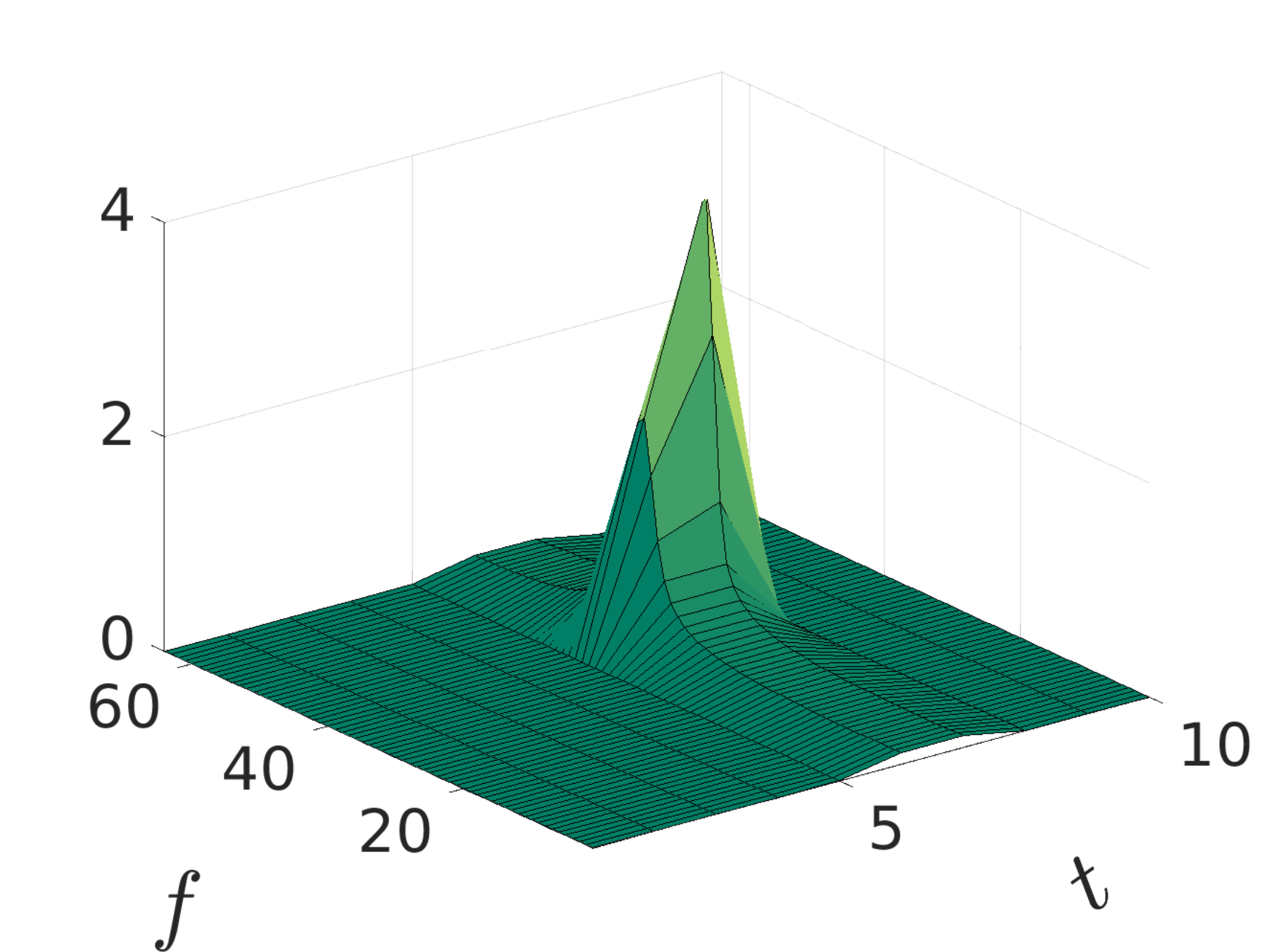}
  \caption{$\phi_1(t,f)$ } 
\end{subfigure}
\quad 
\begin{subfigure}{.22\textwidth}
  \centering
  \includegraphics[width=1\linewidth]{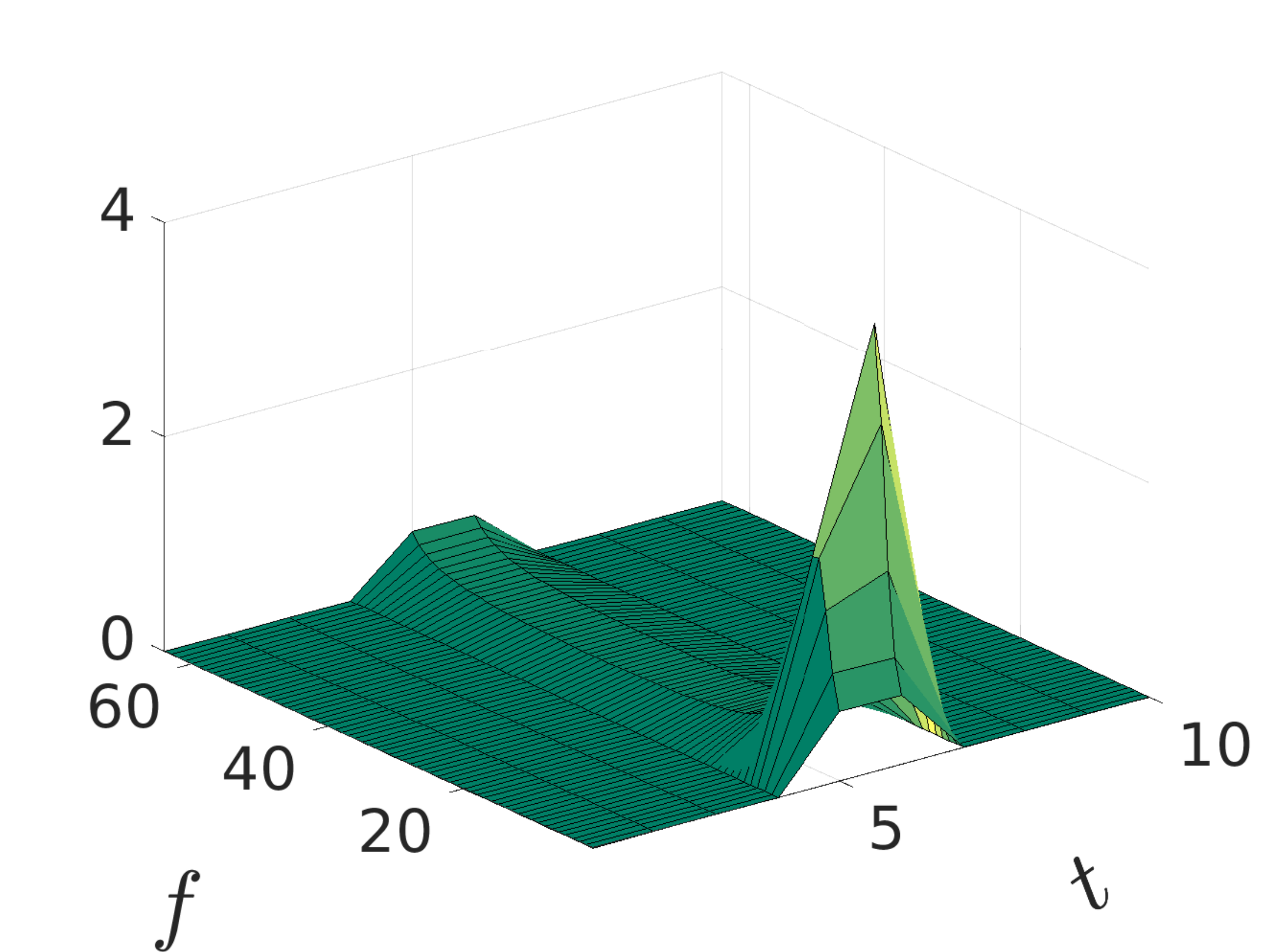}
  \caption{$\phi_2(t,f)$} 
\end{subfigure}\\
\begin{subfigure}{.22\textwidth}
  \centering
  \includegraphics[width=1\linewidth]{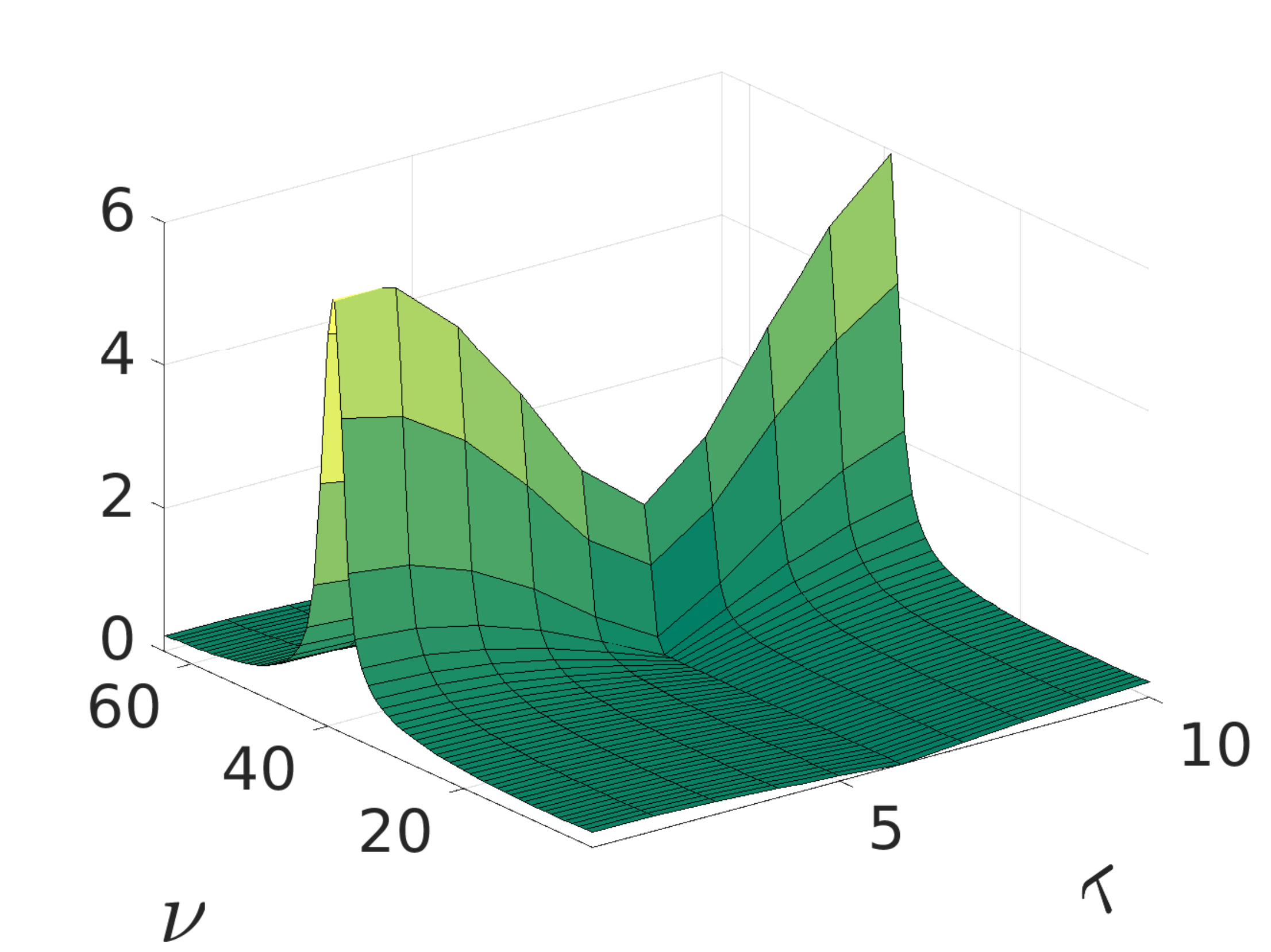}
  \caption{$\phi_1(\tau,\nu)$ } 
\end{subfigure}
\begin{subfigure}{.22\textwidth}
  \centering
  \includegraphics[width=1\linewidth]{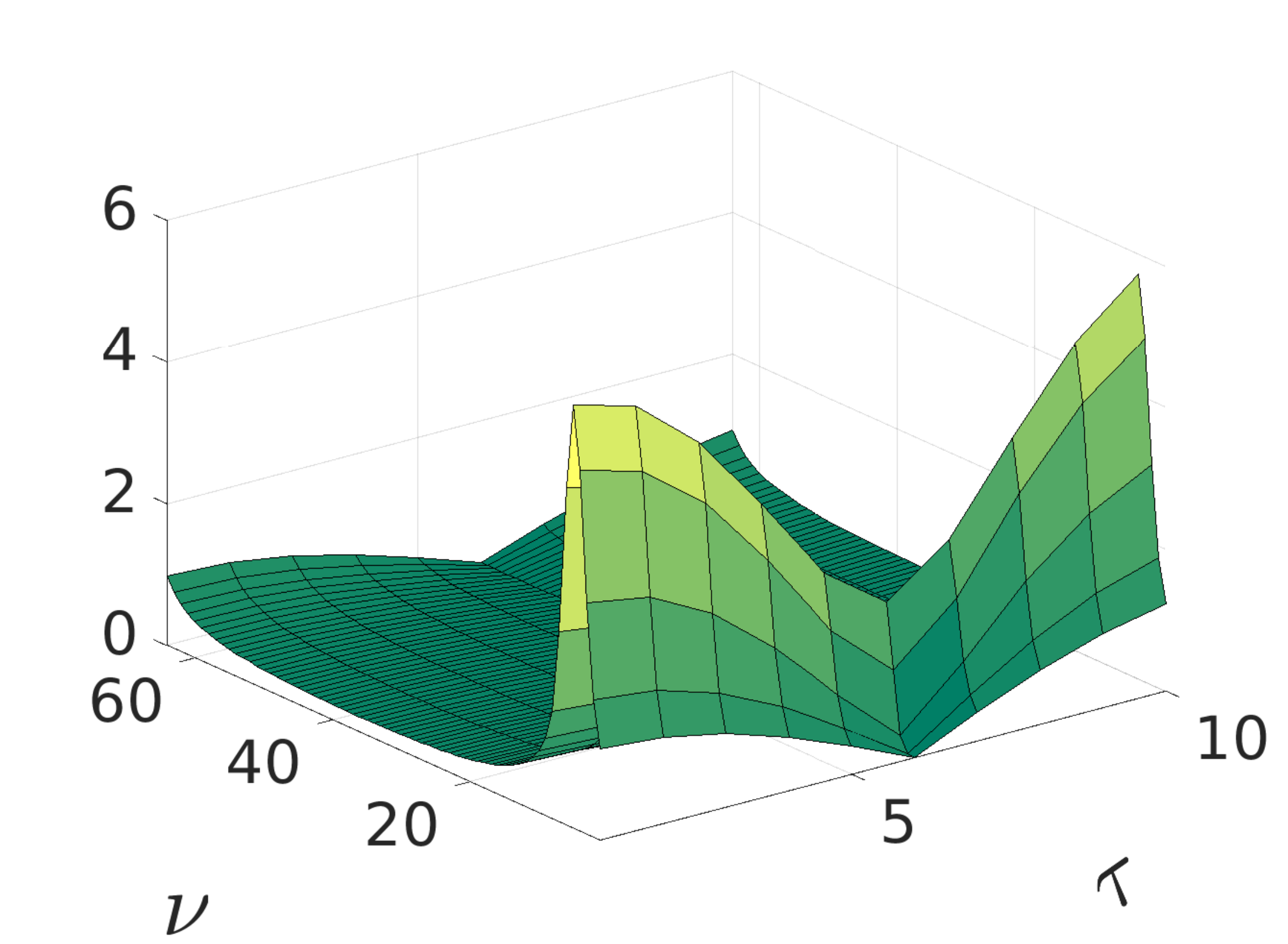}
  \caption{$\phi_2(\tau,\nu)$ } 
\end{subfigure}
  \caption{Eigenwaves in time-frequency, $\phi_n(t,f)$ and delay-Doppler domain, $\phi_n(\tau,\nu)$}
  \label{fig:eigenwave}
  \vspace{-15pt}
\end{figure}

\begin{figure*}
\begin{subfigure}{.24\textwidth}
  \centering
  \includegraphics[width=\linewidth]{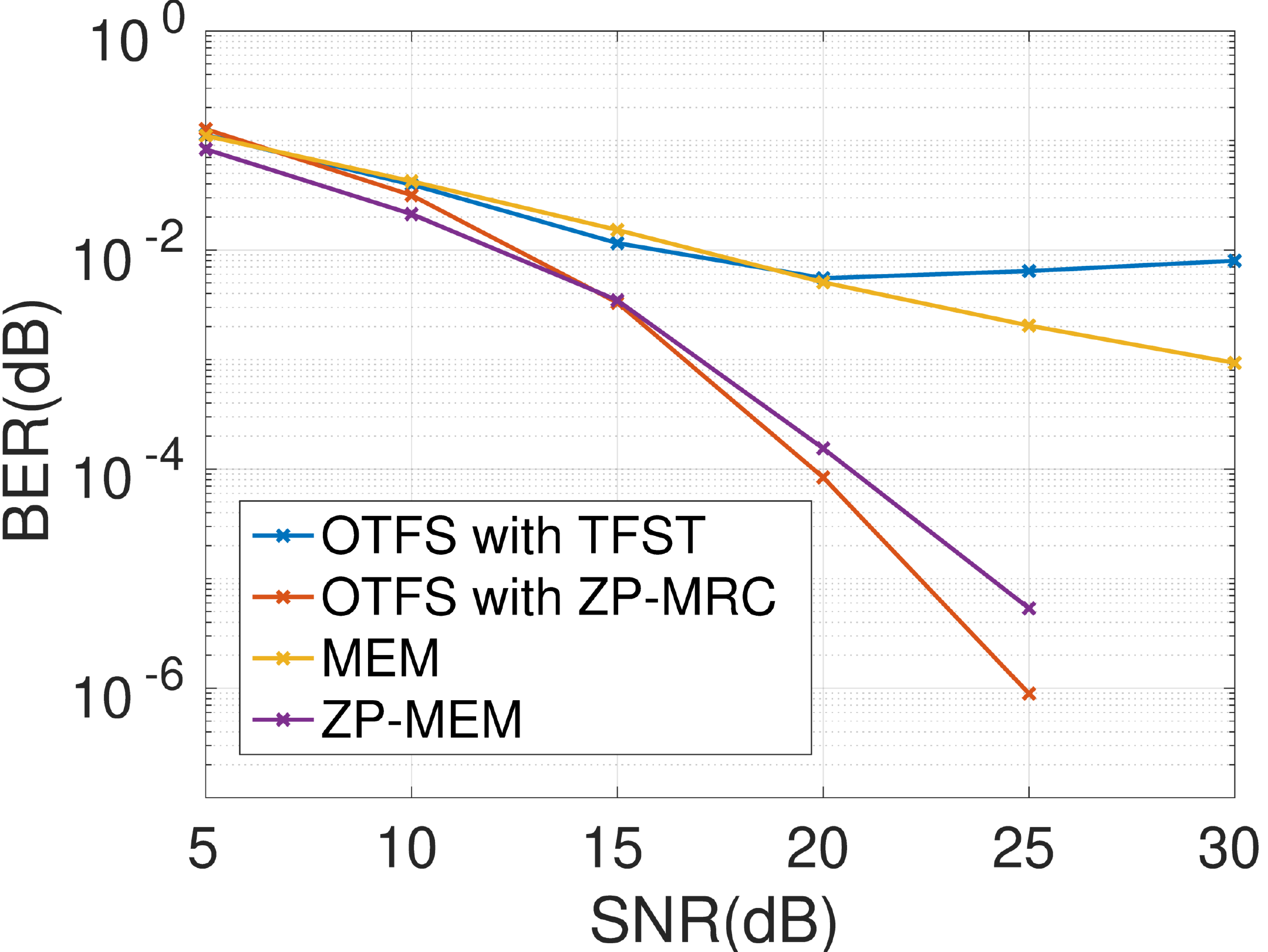}
  \caption{Channel A - BER
  } 
  \label{fig:ber_a}
    \end{subfigure}
    \begin{subfigure}{.235\textwidth}
  \centering
  \includegraphics[width=1\linewidth]{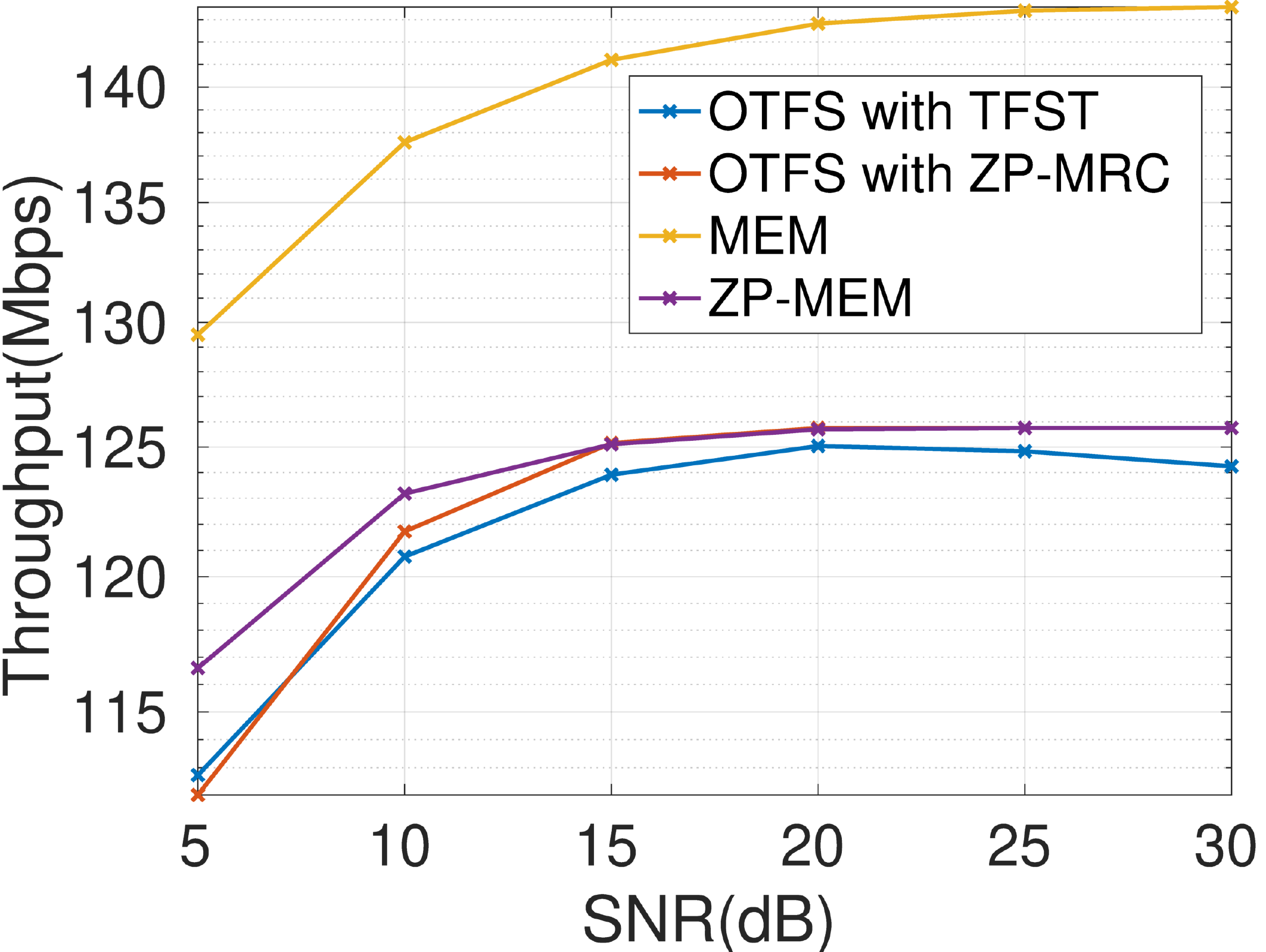}
 \caption{Channel A - Throughput}
  \label{fig:th_a}
\end{subfigure}
\begin{subfigure}{.24\textwidth}
  \centering
  \includegraphics[width=1\linewidth]{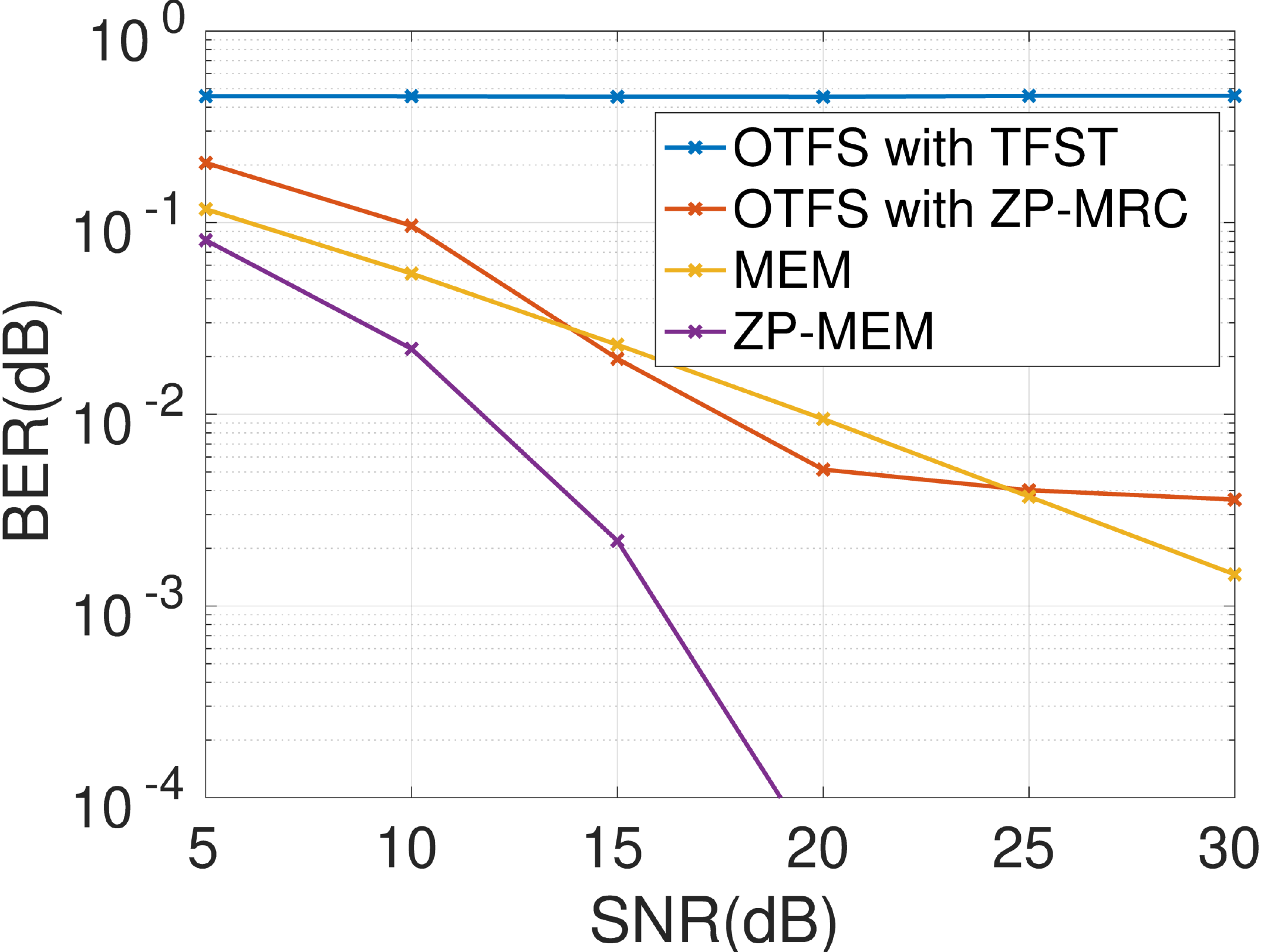}
  \caption{Channel B - BER
  } 
  \label{fig:ber_b}
\end{subfigure}
\begin{subfigure}{.23\textwidth}
  \centering
  \includegraphics[width=1\linewidth]{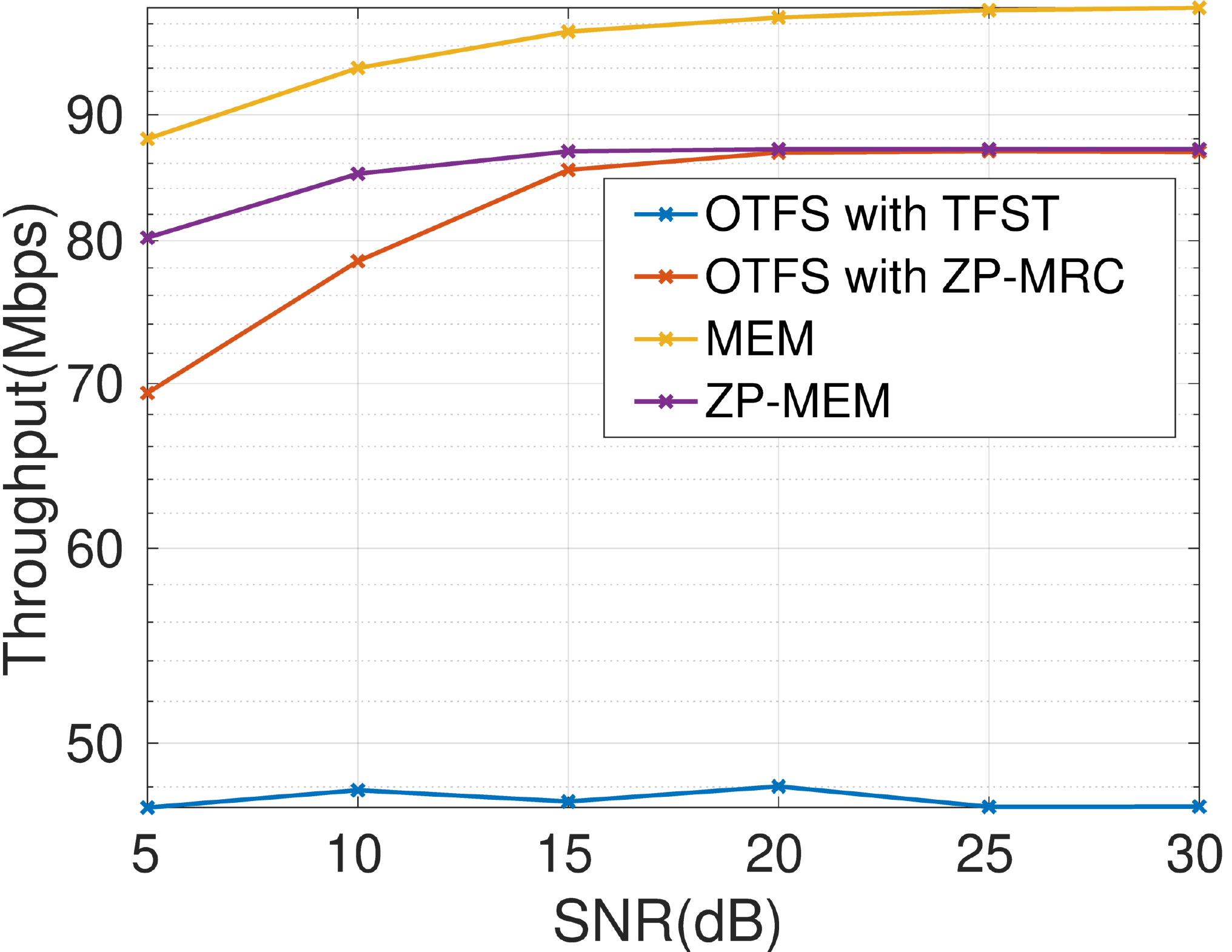}
 \caption{Channel B - Throughput}
  \label{fig:th_b}
\end{subfigure}
  \caption{BER and Throughput comparison between MEM and OTFS for Channel-A and Channel-B with QPSK modulation}
  \label{fig:cb}
  \vspace{-10pt}
\end{figure*}

We analyze the accuracy of MEM modulations without supplemental detectors and present comparisons to OTFS with SToA detectors for non-stationary channels exhibiting varying degrees of non-stationarity. We then demonstrate the generality of our approach for higher-dimensional channels by directly applying MEM modulation to MU-MIMO channels without the need for precoding techniques. In all simulations, we assume perfect CSI at both the transmitter and receiver.

\noindent
\textbf{Non-stationary channels}: We establish the channels in Matlab using the Extended Vehicular A (EVA) model with parameters provided in Table~\ref{tab:parameters}. 
To demonstrate the effects of non-stationarity, we compared our methods with OTFS for two channels: 1) Channel-A, where the resolution of time evolution is one symbol, and 2) Channel-B, where the resolution of time evolution is one subcarrier. In Channel-A and Channel-B, we generate the delay-Doppler response per symbol and per subcarrier, respectively, which also correspond to the stationarity intervals of the two channels. 
OTFS is equipped with the time-frequency single tap (TFST)~\cite{Hong2022DD} detector and Zero-Padded maximal ratio
combining (ZP-MRC)~\cite{thaj2020low} detector, respectively. For a fair comparison, we also implement a Zero-Padded MEM (ZP-MEM) version, where \textit{zero pad} is placed on eigenfunctions with the lowest $\sigma_n$. The ZP length is $1/8$ symbol for both ZP-MEM and ZP-MRC. 

\begin{table}[b]
\vspace{-15pt}
\caption{Parameters of Channel-A and Channel-B}
\renewcommand*{\arraystretch}{1.2}
\centering
\begin{tabular}{|c|c|}
\midrule
\textbf{Parameter} & \textbf{Value} \\ \midrule
Channel model & EVA model \\\midrule 
Bandwidth & Bw $ = 960$ KHz     \\\midrule
Center frequency & $f_c = 5$ GHz \\ \midrule
Subcarriers & $N_s = 64$ subcarriers \\ \midrule
Carrier spacing & $\Delta f = 15$ KHz \\ \midrule
Speed range& $v \in [100, 150]$ km/h \\ \midrule
Symbols per frame & $L_F = 10$ symbols \\ \midrule
Frame per packet & $L_P = 100$ frames \\ \midrule
Stationarity interval & Channel A: $1$ symbol;  Channel B: $1$ subcarrier  \\ \midrule
\end{tabular}
\label{tab:parameters}
\end{table}

To illustrate the geometry of the eigenspace, Figure~\ref{fig:eigenwave} shows an example of \textit{two} time-frequency eigenwaves extracted from the kernel $k(t,f;t',f')$ by HOGMT and their representations in the delay-Doppler Domain. Unlike OFDM and OTFS, the eigenwave is an orthonormal surface across its degrees of freedom instead of an unit division in the time-frequency or the delay-Doppler domain. However, from another perspective, consider a Hilbert space, $\mathbb{H}_{\Phi}$ with basis $\{\phi_n\}$, then each eigenwave can be seen as an unit division in $\mathbb{H}_{\Phi}$. It means MEM analyzes the channel as one unified space (eigenspace) instead of multiple subspaces of its degrees of freedom.

Figure~\ref{fig:ber_a} compares the BER of MEM, ZP-MEM, OTFS with TFST and OTFS with ZP-MRC. MEM has lower BER than OTFS with TFST after 20 dB SNR, but higher BER than both ZP-MEM and OTFS with ZP-MRC. This is because demodulating data symbols on carriers (eigenwaves) with least $\sigma_n$ will enhance the noise as well. ZP-MEM doesn't put data symbols on those eigenwaves, thereby achieving lower BER. On the other hand, ZP-MRC detector can cancel interference among OTFS symbols and thus has the similar BER with ZP-MEM. However, as shown in figure~\ref{fig:th_a}, MEM has the highest throughput due to no zero pad.  

Figure~\ref{fig:ber_b} shows the BER for Channel-B, where the stationarity interval is just one subcarrier. TFST detector doesn't work at all in this case and ZP-MRC detector has a similar BER as MEM because there are more interference at delay-Doppler domain in this channel. Both ZP-MEM and MEM are not affected because interference at delay-Doppler domain would not affect the orthogonality among eigenwaves. MEM still has the highest throughput as shown in figure~\ref{fig:th_b}, while TFST performs much worse in this scenario.  

\noindent
\textbf{Higher dimension channels:} We also validate the generality of MEM to higher dimension by incorporating space domain using 3GPP 38.901 UMa NLOS senario
built on QuaDriga in Matlab. The channel parameters and results are given in Appendix D~\cite{ICC23appendix}. 

\section{Conclusion}
\label{sec:conclusion}
    In this paper, we show the evolution and limitations of current modulation techniques (OFDM, OTFS) for MU-MIMO non-stationary channels and proposed a novel MEM modulation based on HOGMT decomposition. It is able to achieve orthogonality for non-stationary channels and generalizes to higher dimensions by using multidimensional eigenwaves as carriers, which are jointly orthogonal across its DoF (space, time-frequency and delay-Doppler domains). Therefore MEM modulated symbols that are transmitted over multidimensional channels will remain independent of each other, and thus eliminates multidimensional interference without any additional precoding at the transmitter or detectors at the receiver. Moreover, we demonstrate that achieving the diversity gain in eigenspace is equivalent to achieving the diversity gain in DoF, thus validating the generality of MEM modulation.








\bibliographystyle{IEEEtran}
\bibliography{references}

\begin{thebibliography}{10}
\providecommand{\url}[1]{#1}
\csname url@samestyle\endcsname
\providecommand{\newblock}{\relax}
\providecommand{\bibinfo}[2]{#2}
\providecommand{\BIBentrySTDinterwordspacing}{\spaceskip=0pt\relax}
\providecommand{\BIBentryALTinterwordstretchfactor}{4}
\providecommand{\BIBentryALTinterwordspacing}{\spaceskip=\fontdimen2\font plus
\BIBentryALTinterwordstretchfactor\fontdimen3\font minus
  \fontdimen4\font\relax}
\providecommand{\BIBforeignlanguage}[2]{{%
\expandafter\ifx\csname l@#1\endcsname\relax
\typeout{** WARNING: IEEEtran.bst: No hyphenation pattern has been}%
\typeout{** loaded for the language `#1'. Using the pattern for}%
\typeout{** the default language instead.}%
\else
\language=\csname l@#1\endcsname
\fi
#2}}
\providecommand{\BIBdecl}{\relax}
\BIBdecl

\bibitem{ZouICC22}
Z.~Zou, M.~Careem, A.~Dutta, and N.~Thawdar, ``Unified characterization and
  precoding for non-stationary channels,'' in \emph{ICC 2022 - IEEE
  International Conference on Communications}, 2022, pp. 5140--5146.

\bibitem{ZouTCOM23}
{Z. Zou}, M.~Careem, A.~Dutta, and N.~Thawdar, ``{Joint Spatio-Temporal
  Precoding for Practical Non-Stationary Wireless Channels},'' \emph{IEEE
  Transactions on Communications}, pp. 1--1, 2023.

\bibitem{tse2004}
D.~Tse and P.~Viswanath, \emph{{Fundamentals of Wireless Communication}}.\hskip
  1em plus 0.5em minus 0.4em\relax USA: Cambridge University Press, 2005.

\bibitem{OTFS_2018_Paper}
R.~Hadani, S.~Rakib, S.~Kons, M.~Tsatsanis, A.~Monk, C.~Ibars, J.~Delfeld,
  Y.~Hebron, A.~J. Goldsmith, A.~F. Molisch, and A.~R. Calderbank, ``Orthogonal
  time frequency space modulation,'' \emph{CoRR}, vol. abs/1808.00519, 2018.

\bibitem{Raviteja2018eq_otfs}
P.~Raviteja, K.~T. Phan, Y.~Hong, and E.~Viterbo, ``{Interference Cancellation
  and Iterative Detection for Orthogonal Time Frequency Space Modulation},''
  \emph{IEEE Transactions on Wireless Communications}, vol.~17, no.~10, pp.
  6501--6515, 2018.

\bibitem{Thaj2021RxOTFS}
T.~Thaj, E.~Viterbo, and Y.~Hong, ``{Orthogonal Time Sequency Multiplexing
  Modulation: Analysis and Low-Complexity Receiver Design},'' \emph{IEEE
  Transactions on Wireless Communications}, vol.~20, no.~12, pp. 7842--7855,
  2021.

\bibitem{cao2021low}
B.~Cao, Z.~Xiang, and P.~Ren, ``{Low complexity transmitter precoding for MU
  MIMO-OTFS},'' \emph{Digital Signal Processing}, vol. 115, p. 103083, 2021.

\bibitem{pandey2021low}
B.~C. Pandey, S.~K. Mohammed, P.~Raviteja, Y.~Hong, and E.~Viterbo, ``{Low
  complexity precoding and detection in multi-user massive MIMO OTFS
  downlink},'' \emph{IEEE transactions on vehicular technology}, vol.~70,
  no.~5, pp. 4389--4405, 2021.

\bibitem{Guo2022CSI}
\BIBentryALTinterwordspacing
J.~Guo, C.-K. Wen, S.~Jin, and G.~Y. Li, ``{Overview of Deep Learning-based CSI
  Feedback in Massive MIMO Systems},'' 2022. [Online]. Available:
  \url{https://arxiv.org/abs/2206.14383}
\BIBentrySTDinterwordspacing

\bibitem{MATZ20111}
F.~Hlawatsch and G.~Matz, \emph{Wireless Communications Over Rapidly
  Time-Varying Channels}, 1st~ed.\hskip 1em plus 0.5em minus 0.4em\relax USA:
  Academic Press, Inc., 2011.

\bibitem{ICC23appendix}
\BIBentryALTinterwordspacing
Z.~Zou and A.~Dutta, ``{Proofs and Supplementary Material: Multidimensional
  Eigenwave Multiplexing Modulation for Non-Stationary Channels}.'' [Online].
  Available:
  \url{https://www.dropbox.com/s/mjlowl23ciav03x/Appendix_MEM.pdf?dl=0}
\BIBentrySTDinterwordspacing

\bibitem{Bello1963Chara}
P.~Bello, ``{Characterization of Randomly Time-Variant Linear Channels},''
  \emph{IEEE Transactions on Communications Systems}, vol.~11, no.~4, pp.
  360--393, 1963.

\bibitem{1909Mercer}
\BIBentryALTinterwordspacing
J.~Mercer, ``{Functions of Positive and Negative Type, and their Connection
  with the Theory of Integral Equations},'' \emph{Philosophical Transactions of
  the Royal Society of London. Series A, Containing Papers of a Mathematical or
  Physical Character}, vol. 209, pp. 415--446, 1909. [Online]. Available:
  \url{http://www.jstor.org/stable/91043}
\BIBentrySTDinterwordspacing

\bibitem{ali2017beamforming}
E.~Ali, M.~Ismail, R.~Nordin, and N.~F. Abdulah, ``{Beamforming techniques for
  massive MIMO systems in 5G: overview, classification, and trends for future
  research},'' \emph{Frontiers of Information Technology \& Electronic
  Engineering}, vol.~18, no.~6, pp. 753--772, 2017.

\bibitem{Hong2022DD}
Y.~Hong, T.~Thaj, and E.~Viterbo, \emph{{Delay-Doppler Communications:
  Principles and Applications}}, 1st~ed.\hskip 1em plus 0.5em minus 0.4em\relax
  USA: Elsevier Science., 2022.

\bibitem{thaj2020low}
T.~Thaj and E.~Viterbo, ``{Low complexity iterative rake decision feedback
  equalizer for zero-padded OTFS systems},'' \emph{IEEE transactions on
  vehicular technology}, vol.~69, no.~12, pp. 15\,606--15\,622, 2020.

\end{thebibliography}




\end{document}


\def\eg{\mbox{\em e.g.}, }

\title{Proofs and Supplementary Material:\\ Unified Characterization and Precoding for Non-Stationary Channels}

\author{
\begin{tabular}[t]{c@{\extracolsep{8em}}c} 
Zhibin Zou, Maqsood Careem, Aveek Dutta & Ngwe Thawdar \\
Department of Electrical and Computer Engineering & US Air Force Research Laboratory \\ 
University at Albany SUNY, Albany, NY 12222 USA & Rome, NY, USA \\
\{{zzou2, mabdulcareem, adutta\}@albany.edu} & 
ngwe.thawdar@us.af.mil
\end{tabular}
\vspace{-3ex}
}


    
\maketitle









\setcounter{equation}{34}
\input{appendix_NLP}

\bibliographystyle{IEEEtran}
\bibliography{references, ref_precoding}